\def\keV{{\rm\thinspace keV}}

\documentclass[usegraphicx,usenatbib]{mn2e}

\usepackage{fixltx2e}
\usepackage{mathptmx}

\include{defn}
\begin{document}

\title[A Spin on Bondi Flow ]{Bondi flow from a slowly rotating hot atmosphere}

\author[Narayan \& Fabian]
{\parbox[]{6.5in}{{Ramesh Narayan$^1\thanks{E-mail:rnarayan@cfa.harvard.edu}$ and Andrew C. Fabian$^2\thanks{E-mail:acf@ast.cam.ac.uk}$}\\
\footnotesize
$^1$Harvard-Smithsonian Center for Astrophysics, 60 Garden Street,
Cambridge, MA 02138, USA\\
$^2$Institute of Astronomy, Madingley Road, Cambridge CB3 0HA\\}}

\maketitle
  
\begin{abstract} A supermassive black hole in the nucleus of an
  elliptical galaxy at the centre of a cool-core group or cluster of
  galaxies is immersed in hot gas. Bondi accretion should occur at a
  rate determined by the properties of the gas at the Bondi radius and
  the mass of the black hole. X-ray observations of massive nearby
  elliptical galaxies, including M87 in the Virgo cluster, indicate a
  Bondi accretion rate $\dot{M}_{\rm B}$ which roughly matches the
  total kinetic power of the jets, suggesting that there is a tight
  coupling between the jet power and the mass accretion rate.  While
  the Bondi model considers non-rotating gas, it is likely that the
  external gas has some angular momentum, which previous studies have
  shown could decrease the accretion rate drastically. We investigate
  here the possibility that viscosity acts at all radii to transport
  angular momentum outward so that the accretion inflow proceeds
  rapidly and steadily. The situation corresponds to a giant Advection
  Dominated Accretion Flow (ADAF) which extends from beyond the Bondi
  radius down to the black hole. We find solutions of the ADAF
  equations in which the gas accretes at just a factor of a few less
  than $\dot M_{\rm B}$. These solutions assume that the atmosphere
  beyond the Bondi radius rotates with a sub-Keplerian velocity and
  that the viscosity parameter is large, $\alpha\geq 0.1$, both of
  which are reasonable for the problem at hand.  The infall time of
  the ADAF solutions is no more than a few times the free-fall
  time. Thus the accretion rate at the black hole is closely coupled
  to the surrounding gas, enabling tight feedback to occur. We show
  that jet powers of a few per cent of $\dot M_{\rm B}c^2$ are
  expected if either a fraction of the accretion power is channeled
  into the jet or the black hole spin energy is tapped by a strong
  magnetic field pressed against the black hole by the pressure of the
  accretion flow. We discuss the Bernouilli parameter of the flow, the
  role of convection, and the possibility that these as well as MHD
  effects may invalidate the model. If the latter comes to pass, it
  would imply that the rough agreement between observed jet powers and
  the Bondi accretion rate is a coincidence and jet power is
  determined by factors other than the mass accretion rate.
\end{abstract}

\begin{keywords}
  X-rays: galaxies --- galaxies: clusters ---
  intergalactic medium: accretion, accretion discs --- black hole physics
\end{keywords}

\section{Introduction}\label{intro}

The nuclei of massive elliptical galaxies at the centres of cool-core
groups and clusters of galaxies have powerful relativistic jets which
inject energy into the surrounding hot gas. This prevents the
intracluster gas from radiatively cooling and collapsing onto the
galaxy, thus stifling its growth (McNamara \& Nulsen 2007 and
references therein). Such feedback is now a common ingredient in our
understanding of the evolution of massive galaxies (Croton et al.
2006; Hopkins et al. 2006).

The mode of fuelling of the massive black hole at the galaxy nucleus,
which energises the jets, is unclear. Bondi (1952) accretion is often
invoked since the black hole is sitting in the densest part of the hot
cluster (or group) atmosphere. Observations of the gas around the
Bondi radius in M87 indicate that Bondi accretion may indeed provide a
suitable mass supply rate (di Matteo et al. 2003). Others argue that
it cannot provide enough fuel to power more powerful, distant objects
(Rafferty et al. 2006), and that cold gas clouds may instead be
required (Pizzolato \& Soker 2005). Regardless, the Bondi model
considers gas with vanishing angular momentum, whereas in a realistic
situation the incoming gas is likely to have non-negligible rotation.
Hence, it is not clear that the Bondi accretion rate $\dot{M}_{\rm B}$
is at all relevant.

Most nuclei in the centres of cool core clusters show no sign of a
dense, radiatively-efficient accretion disc. Some of the most powerful
ones do not even show any detectable X-ray point source
(Hlavecek-Larrondo \& Fabian 2011), which is difficult to explain in
cold mode accretion. In the case of M87, there is clear evidence that
both the accretion flow and the jets themselves are radiatively
inefficient (di Matteo et al. 2003). This indicates that the flow must
be advection dominated, i.e., the gravitational energy released in the
flow must be carried into the centre rather than radiated locally (see
Narayan, Mahadevan \& Quataert 1998; Kato, Fukue \& Mineshige 2008;
Narayan \& McClintock 2008; for reviews of advection-dominated
accretion flows, or ADAFs). A fraction of the energy must then be
efficiently transferred to the jets once the accreting gas reaches the
centre.

The above conclusion is supported by a study of 8 other massive nearby
elliptical galaxies where the gas properties close to the Bondi radius
can be observed or reasonably extrapolated (Allen et al. 2006). In all
these cases, the Bondi mass accretion rate $\dot{M}_{\rm B}$
determined at the Bondi radius $r_{\rm B}$ correlates well with the
power of the jets $P_{\rm j}$, where the latter is measured from the
bubbles inflated by the jets in the surrounding gas. Writing the jet
power as $P_{\rm j}=\eta_{\rm j}\dot M_{\rm B}c^2$, the jet production
efficiency factor $\eta_{\rm j}$ is found to be about 2 per cent. This
is a rather large efficiency and underscores the need for mass at
approximately the Bondi accretion rate reaching the gravitational
radius of the black hole $r_{\rm g}$. There is little room for any
inefficiency in the transport of mass to the centre, e.g., through
mass loss in outflows along the way.

We are concerned here whether an ADAF can be established in galactic
nuclei and whether the mass accretion rate is comparable to the Bondi
rate.  The range of jet power in the systems discussed above is
between $10^{43} - 10^{45}~{\rm erg\,s^{-1}}$, so the Eddington ratio
(power emitted to Eddington limit) is $10^{-4} - 10^{-2}$ for a black
hole of mass $10^9 M_\odot$, and ten times less for
$10^{10}M_\odot$. This is very much in the regime where an ADAF is
expected (Narayan \& Yi 1995b; Narayan \& McClintock 2008).  Moreover,
as noted by Narayan \& Yi (1994, 1995a) and Fabian \& Rees (1995), and
confirmed in more recent investigations (Narayan \& McClintock 2008),
the large thermal pressure of an ADAF may be especially good for the
production and collimation of jets. Thus, it is natural to consider an
ADAF-like accretion model for systems with powerful jets.

ADAFs have been well studied since the work of Narayan \& Yi (1994,
1995ab) and Abramowicz et al. (1995). However, in much of the previous
work, the outer edge of the solution was generally taken to be either
of a self-similar form (e.g., Chen, Abramowicz \& Lasota 1997; Popham
\& Gammie 1998) or a geometrically thin disk that evaporates to form
the ADAF (e.g., Narayan, Kato \& Honma 1997; Manmoto et
al. 2000). Neither of these boundary conditions is relevant for
understanding accretion from an external medium.  Since an ADAF is
essentially space filling, we expect the accretion flow to match
smoothly on to the external medium without any shocks or other kinds
of discontinuities. We investigate in this paper exactly how this
matching occurs when we have a slowly rotating external medium
(eq. \ref{rotpar} gives a quantitative measure of what we mean by slow
rotation).

Previous studies of Bondi-like accretion with angular momentum have
generally considered inviscid flows.  Proga \& Begelman (2003)
carried out two-dimensional axisymmetric simulations and showed that
an equatorial torus forms because of the angular momentum barrier and
that this torus constrains the amount of polar accretion.  Krumholz,
McKee \& Klein (2005) extended their work and developed approximate
formulae for the mass accretion rate as a function of the vorticity of
the external gas, and Cuadra et al. (2006) carried out detailed
simulations of inviscid accretion on to Sagittarius A$^*$ at the
Galactic Centre.  Recently, Inogamov \& Sunyaev (2010) proposed an
accretion model for M87. As in the other studies cited here, the
centrifugal barrier causes the inviscid accreting gas to form a torus
well inside $r_{\rm B}$. Inogamov \& Sunyaev assume that viscosity
then turns on at smaller radii and suggest that the torus will thus
feed a standard thin accretion disc on the inside, which might
evaporate into an ADAF at yet smaller radii.  The presence of the thin
disk segment in their model causes the total inflow time of the gas
from the Bondi radius $r_{\rm B}$ to the black hole gravitational
radius $r_g$ to be far longer than for a Bondi flow or (as we shall
see) an ADAF. Self-adjustment of the feedback, in which the jet power
responds to conditions (e.g. cooling time) beyond $r_{\rm B}$, then
becomes very difficult, with large hysteresis expected.

In contrast to the above studies, we are interested in viscous
accretion. The closest paper to our work is Park (2009).  For
technical reasons, that work focused on extremely hot external media
($T_{\rm ext} > 10^{9}$\,K) for which the Bondi radius is much closer
to the black hole than in real systems. We consider more realistic
external conditions ($T_{\rm ext} \sim 10^{6-7}$\,K). We also study in
more detail the transition from a Bondi flow to an ADAF as the
external rotation is varied.

As in Park (2009), we require the flow to be continuous out through
$r_{\rm B}$ and beyond. Such a model ensures that the accretion power
is as well coupled with the conditions in the outer gas as possible,
thereby allowing for the most efficient feedback.  We moreover require
that outflows, and significant radial exchanges of energy within the
ADAF, are suppressed.  We postulate that relativistic jets are created
and mechanically powered very efficiently (but very radiatively
inefficiently) by the accreting gas close to the black hole, but how
this occurs is beyond the scope of the present work. We limit
ourselves to a more basic question: Can an idealised ADAF transfer a
high enough mass accretion rate from beyond $r_{\rm B}$ down to
$r_{\rm g}$?
  
\section{Spherical ADAF Model}\label{ADAF}

\subsection{Viscous Accretion Flow: Conservation Laws}\label{conslaws}

Since we are primarily interested in slowly-rotating, steady, viscous
accretion flows, we assume that the density and pressure of the gas
are distributed spherically at each radius. We also assume that all
quantities are independent of time (steady state assumption).  We thus
focus only on radial variations. Under these assumptions, the mass
accretion rate $\dot{M}$ at radius $r$ is given by
\begin{equation}\label{mdot}
\dot{M} = -4\pi r^2\rho v = {\rm constant},\label{mass}
\end{equation}
where $\rho(r)$ is the density and $v(r)$ is the radial velocity; the
latter is taken to be negative when gas flows inwards. When
considering accretion flows in which rotational support is important,
e.g., geometrically thin disks, or ADAFs with more rotation than we
consider here, the factor $4\pi r^2$ in the above relation is replaced
by $(2\pi r)(2H)$, where $H(r)$ is the ``vertical'' scale height of
the gas at radius $r$. In the simpler approximation considered here,
we effectively set $H=r$, which could be interpreted as a
geometrically very thick disk. Except for this difference, the
equations we consider are identical to those described in Narayan et
al. (1997).

To mock up relativistic gravity in our Newtonian model, we assume a
gravitational potential (Paczy\'nski \& Wiita 1980)
\begin{equation}\label{pw}
\phi(r) = -\frac{GM}{(r-r_g)}, \qquad r_g=\frac{2GM}{c^2},
\end{equation}
where $M$ is the mass of the central black hole. Correspondingly, the
Keplerian angular frequency $\Omega_K$ is given by
\begin{equation}
\Omega_K^2 = \frac{GM}{(r-r_g)^2r}.
\end{equation}
Making the replacement $p=\rho c_s^2$, where $c_s$ is the (isothermal)
sound speed, we write the steady state radial momentum equation as
\begin{equation}\label{mmtm}
v\,\frac{dv}{dr} = -(\Omega_K^2-\Omega^2)r -\frac{1}{\rho}\,\frac{d}{dr}
(\rho c_s^2),
\end{equation}
where $\Omega$ is the angular velocity of the gas on the equatorial
plane. Our spherical model is most accurate when the centrifugal
acceleration on the gas is much weaker than the gravitational
acceleration; this corresponds to the condition $\Omega^2 \ll
\Omega_K^2$.

We model viscosity via the standard $\alpha$-prescription (Shakura \&
Sunyaev 1973) in which the kinematic coefficient of viscosity $\nu$ is
written as
\begin{equation}
\nu=\alpha c_s H = \alpha c_s r,\label{visc}
\end{equation}
with $\alpha$ taken to be a constant. However, we do not set the shear
stress equal to $\alpha p$, but use a more physical prescription in
which the stress is proportional to the angular velocity gradient:
\begin{equation}
{\rm shear~stress} \equiv \sigma_{r\phi} = \nu \rho r d\Omega/dr.  \label{shearstress}
\end{equation}
The angular momentum equation
then takes the form (Narayan et al. 1997)
\begin{equation}\label{angmmtm0}
v\frac{d}{dr}(\Omega r^2) = \frac{1}{\rho r^2}\frac{d}{dr}
\left(\alpha\rho c_s r^5\frac{d\Omega}{dr}\right),
\end{equation}
which on integration gives
\begin{equation}\label{angmmtm}
\frac{d\Omega}{dr} = \frac{v(\Omega r^2-j)}{\alpha r^3c_s}.
\end{equation}
The quantity $j$ is an integration constant with dimensions of
specific angular momentum.

Finally, energy conservation gives
\begin{equation}\label{energy}
\frac{\rho v}{(\gamma-1)}\frac{dc_s^2}{dr} - c_s^2v\frac{d\rho}{dr} =
\alpha \rho c_s r^3 \left(\frac{d\Omega}{dr}\right)^2,
\end{equation}
where $\gamma$ is the adiabatic index of the gas, which is set to 5/3
for all the numerical models presented here.  The left-hand side of
equation (\ref{energy}) represents the Lagrangian time derivative of
the entropy of the gas. This term is usually referred to as the energy
advection term. The term on the right-hand side of the equation
describes the heating rate due to viscous dissipation. In the spirit
of a radiatively inefficient flow (ADAF), we ignore radiative cooling
altogether, Thus, we set advection equal to heating to obtain equation
(\ref{energy}).

We should note the following inconsistency in the above
equations\footnote{We thank the referee for pointing this out to us}.
While we have included the effect of viscosity through the shear
stress in the angular momentum equation (\ref{angmmtm0}), we have
neglected corresponding terms in the radial momentum equation
(\ref{mmtm}).  Under the assumptions of our model (pure radial flow,
no gradients in the transverse direction, etc.), the $rr$ component of
the stress takes the form (Landau \& Lifshitz 1959)
\begin{equation}
\sigma_{rr} = -\rho c_s^2 +\frac{4}{3}\eta\rho r\frac{d(v_r/r)}{dr}
+\xi\rho\frac{1}{r^2}\frac{d(r^2 v_r)}{dr}, \label{sigmarr}
\end{equation}
where $\xi$ is the kinematic bulk viscosity.  The last term in
equation (\ref{mmtm}) should thus be written as
$(1/\rho)\,d(\sigma_{rr})/dr$ with the above form of $\sigma_{rr}$,
not just as $-(1/\rho)\,d(\rho c_s^2)/dr$.

Traditionally, in accretion disk models, the viscous terms in
$\sigma_{rr}$ are neglected on the grounds that $v_r$ is much smaller
than $c_s$ and so these terms are small compared to the pressure.
This is no longer obvious for the slowly-rotating solutions presented
here, for which $v_r$ is fairly large.  Nevertheless, we make this
assumption for easy comparison with previous work.  A major goal of
the present work is to study the transition from the rapidly-rotating
ADAF regime to the non-rotating Bondi regime.  The viscous terms in
$\sigma_{rr}$ survive even for pure radial flow and ought to be
included in a self-consistent model of spherical inflow.  Since these
terms are neglected in the Bondi model, in the same spirit we neglect
them in our model as well. We leave for the future an investigation of
the effect of these terms on both the Bondi solution and our
slowly-rotating solution.

\subsection{The Inner Supersonic Region}\label{supersonic}

The equations in \S\ref{conslaws} correspond to a viscous rotating
accretion flow. Once the accreting gas passes inside the sonic radius
$r_s$ and becomes supersonic, we expect viscosity to be much reduced
and perhaps even to vanish (Narayan 1992; Kato \& Inagaki 1994; Kley
\& Papaloizou 1997).  For this region of the flow, we simplify the
equations by setting $\alpha=0$, thus dropping all terms related to
viscosity. From the angular momentum equation (\ref{angmmtm0}) we see
that the specific angular momentum is then a constant:
\begin{equation}
r<r_s: \quad \ell_{\rm in} \equiv \Omega r^2 = {\rm constant}.
\end{equation}
Similary, from the energy equation (\ref{energy}), we see that the
entropy of the gas is constant:
\begin{equation}
r<r_s: \quad s_{\rm in} \equiv \frac{c_s^2}{\rho^{(\gamma-1)}} = {\rm
constant}
\end{equation}
Finally, by combining the various conservation laws, we can show that
the Bernoulli parameter $\cal B$ of the gas is also constant. This
gives the condition
\begin{eqnarray}
r<r_s: \quad {\cal B} &\equiv & \frac{v^2}{2}+\frac{\ell_{\rm
in}^2}{2r^2} -\frac{GM}{(r-r_g)} + \nonumber\\ 
&&\frac{\gamma\, s_{\rm
in}}{(\gamma-1)\,r^{2(\gamma-1)} \,|v|^{(\gamma-1)}} 
\left(\frac{\dot{M}}{4\pi}\right)^{(\gamma-1)} \nonumber\\ &=& {\rm constant}.
\label{bernoulli}
\end{eqnarray}
Using the final relation, along with the values of the conserved
quantities $\dot{M}$, $\ell_{\rm in}$, $s_{\rm in}$ and $\cal B$, we
can solve for the radial velocity $v$ as a function of $r$ in the
supersonic region. This immediately gives all the other quantities.

\subsection{Boundary Conditions}\label{bcs}

Our model accretion flow consists of two regions: a viscous subsonic
region which extends from the sonic radius $r_s$ out to some large
outer radius $r_{\rm out}$, and an inviscid supersonic region which
extends from the sonic radius down to the black hole. Finding the
solution in the viscous region requires solving a boundary value
problem involving a number of differential equations.\footnote{Once we
have the solution in the viscous subsonic region, we can compute the
values of $\dot{M}$, $\ell_{\rm in}$, $s_{\rm in}$ and $\cal B$ at
$r_s$. We can then directly calculate the solution in the supersonic
region. The latter involves only algebraic equations (see
\S\ref{supersonic}), not differential equations.}  Equations
(\ref{mmtm}), (\ref{angmmtm}) and (\ref{energy}) represent three first
order ordinary differential equations, which require three boundary
conditions. In addition, the constants $\dot{M}$ and $j$ are
eigenvalues, which require two more boundary conditions. Finally, the
location of the sonic radius $r_s$ has to be determined as part of the
solution, so this requires yet another boundary condition.  Thus we
need to supply a total of six boundary conditions.

The three differential equations (\ref{mmtm}), (\ref{angmmtm}) and
(\ref{energy}), in combination with equation (\ref{mdot}), can be
reduced to the following relation:
\begin{equation}\label{sonic}
(\gamma c_s^2-v^2)\frac{d\ln |v|}{dr} = (\Omega_K^2-\Omega^2)r
-\frac{2\gamma c_s^2}{r} + \frac{(\gamma-1)(\Omega r^2-j)^2v}
{\alpha r^3c_s},
\end{equation}
which becomes singular when $\gamma c_s^2-v^2=0$.  The radius at which
this happens is the sonic radius $r_s$, where the flow speed $|v|$ is
equal to the adiabatic sound speed $c_s\sqrt{\gamma}$.  In order to
have a smooth flow through $r_s$, the quantity on the right hand side
of equation (\ref{sonic}) should vanish. We thus obtain the following
two boundary conditions:
\begin{eqnarray}
r=r_s &:&\quad \gamma c_s^2-v^2=0, \label{sonic1} \\
r=r_s &:& \quad (\Omega_K^2-\Omega^2)r_s
-\frac{2\gamma c_s^2}{r_s} + \frac{(\gamma-1)(\Omega r_s^2-j)^2v}
{\alpha r_s^3c_s}=0. \label{sonic2}
\end{eqnarray}

Viscous accretion flows have another boundary condition on the inside,
which is usually applied as a no-torque condition at some
radius.\footnote{This boundary condition is needed only for the more
physical viscous stress prescription (eq. \ref{shearstress}) used
here. If the shear stress is written in the simpler form $\alpha p$,
there is one fewer differential equation and the additional boundary
condition is not needed (see Narayan et al. 1997 for a discussion). In
fact, since pressure never vanishes in an accretion solution, the
$\alpha p$ stress prescription does not have vanishing stress at any
radius. } In the most elaborate version of the theory, one would apply
the no-torque condition at the black hole horizon ($r=r_g$); however,
this tends to make the numerical computations very difficult.  It also
introduces some subtlety into the problem since the behavior of
viscosity in the supersonic plunging region of the flow is poorly
understood (Narayan 1992; Kato \& Inagaki 1994).  In this paper, we
have assumed for simplicity that viscosity vanishes inside the sonic
radius. One consequence of this approximation is that the specific
angular momentum of the gas $\Omega r^2$ becomes constant in the
plunging region.  Motivated by this fact, we set $d(\Omega r^2)/dr=0$
as a boundary condition on the viscous solution at $r=r_s$. This
ensures a smooth transition across the sonic radius. Making use of
equation (\ref{angmmtm}), the condition can be written as
\begin{equation}\label{bc3}
r=r_s: \quad \Omega r_s^2-j = -\frac{2\alpha c_s\Omega r_s^2}{v}.
\end{equation}

The remaining three boundary conditions are applied at the outer
radius $r_{\rm out}$ of the solution (Yuan 1999). We choose $r_{\rm out}$ to be
large enough that it lies well into the external uniform medium.  In
analogy with the Bondi problem, the temperature of the external gas,
or equivalently the sound speed, and the density of the gas provide
two outer boundary conditions:
\begin{eqnarray}
r=r_{\rm out}&:& \quad c_s = c_{\rm out}, \label{out1} \\
r=r_{\rm out}&:& \quad \rho = \rho_{\rm out}. \label{out2}
\end{eqnarray}
In the numerical solutions presented here, we set $c_{\rm
  out}=10^{-3}c$, which corresponds to a temperature of
$6.5\times10^6$\,K (assuming a mean molecular weight of 0.6), a
reasonable choice for the interstellar medium at the centre of a
galaxy. In the case of the density, we arbitrarily select $\rho_{\rm
  out}=1$.  After the fact, we can rescale the density profile by a
constant factor so as to satisfy the required value of $\rho_{\rm
  out}$.  This approach is allowed by the fact that the equations are
linear in the density.\footnote{This is true only because we have
  ignored all cooling terms. If we include a detailed cooling model,
  the physics will no longer be linear in $\rho$.}

For the third boundary condition, we fix the angular velocity of the
external gas:
\begin{equation}\label{out3}
r=r_{\rm out}: \quad \Omega = \Omega_{\rm out}.
\end{equation}
However, we note the following complication.  Because we are solving
viscous accretion equations with a constant $\alpha$, the solution
naturally tends to a state of rigid rotation on the outside. For radii
outside the Bondi radius,
\begin{equation}\label{rbondi}
r_{\rm B} = \frac{GM}{c_{\rm out}^2}= \frac{1}{2}
\left( \frac{c}{c_{\rm out}}\right)^2 r_g,
\end{equation}
the black hole gravity is too weak to influence the dynamics of the
gas -- pressure is more important here. As a result, viscosity drives
the gas towards $d\Omega/dr=0$.  In a real galactic nucleus, this is
precisely the region where the gravitational potential of the galaxy
will take over and the gas will transition to the rotation curve of
the galaxy (see Quataert \& Narayan 2000 for a study of Bondi
accretion in such a potential). Since we have not included the
galactic contribution to the potential (\ref{pw}), our equations
enforce a rigidly rotating external medium.  The problem with this is
that the centrifugal acceleration $\Omega^2r$ increases without bound
at large radius, which is unphysical. To avoid this problem we choose
$r_{\rm out}$ to be only a factor of a few (not more than 10) larger
than $r_{\rm B}$. By making this choice, we ensure that the
centrifugal acceleration does not become too large on the outside. At
the same time, we make sure that $r_{\rm out}$ is large enough for the
solution to asymptote to the conditions in the external medium.

The parameter $\Omega_{\rm out}$ determines whether the external gas
is rotating slowly or rapidly.  The boundary between the black
hole-dominated accretion flow and the external medium is located at
$r\sim r_{\rm B}$, and the Keplerian angular velocity $\Omega_{K,\rm
B}$ at this radius is given by
\begin{equation}
\Omega_{K,\rm B} = \left(\frac{GM}{r_{\rm B}^3}\right)^{1/2} = 
2\left( \frac{c}{c_{\rm out}}\right)^{-3} \frac{c}{r_g}.
\end{equation}
We thus define the following dimensionless rotation parameter
$\cal{R}$:
\begin{equation}\label{rotpar}
{\cal R} \equiv \frac{\Omega_{\rm out}}{\Omega_{K,\rm B}}
=\frac{1}{2}\Omega_{\rm out}\left(\frac{c}{c_{\rm out}}\right)^3.
\end{equation}
When ${\cal R}\ll1$, we say that the external medium rotates slowly,
whereas as $\cal{R}$ approaches unity, the medium rotates rapidly.  We
are primarily interested in the slowly rotating case.

If the external medium rotates slowly enough, the gas may be able to
accrete directly into the black hole without any need for viscous
transport of angular momentum.  We would then have something very
similar to the Bondi solution. The critical angular momentum of the
external gas at which we expect this transition to take place is the
specific angular momentum of the marginally stable orbit $\ell_{\rm
  ms}$, which for the potential (\ref{pw}) is
\begin{equation}\label{lms}
\ell_{\rm ms} = \sqrt{\frac{27}{8}}\,c r_g.
\end{equation}
Correspondingly, we can express the angular momentum of the external
gas as the following dimensionless ratio
\begin{equation}\label{ellpar}
{\cal L} \equiv \frac{\ell_{\rm out}}{\ell_{\rm ms}}
=\frac{\Omega_{\rm out} r_{\rm B}^2}{\ell_{\rm ms}}
=0.136\, \Omega_{\rm out}\left(\frac{c}{c_{\rm out}}\right)^4.
\end{equation}
When ${\cal L}\gg1$, we expect the flow to be viscously driven and to
resemble an ADAF solution, whereas when ${\cal L}\ll1$, the flow
should be practically identical to the Bondi solution. These
expectations are borne out by the numerical solutions described in
\S\ref{results}. For our choice of $c_{\rm out} = 10^{-3}c$, ${\cal
  L}=1$ corresponds to ${\cal R} = 0.0037$.

\section{Numerical Results}\label{results}

Since the viscous accretion equations tend to be very stiff, we use a
relaxation method (Press et al. 1992) to solve them.\footnote{The
  simpler shooting method is adequate if the outer radius is not too
  large, e.g., $r_{\rm out}/r_g < 10^3$. However, for realistic
  external media with $c_{\rm out}/c\sim 10^{-3}$, we need to
  calculate solutions out to $r_{\rm out}/r_g > 10^6$. In our
  experience, relaxation is the only sure way to obtain such
  solutions.} Figure 1 shows sample solutions corresponding to
$\alpha=0.1$, $\gamma=5/3$, $c_{\rm out} = 10^{-3}$ and $\rho_{\rm
  out}=1$ (the value of $\rho_{\rm out}$ is arbitrary since we can
rescale the density profile to any external density as needed,
\S\ref{bcs}).  Four solutions are shown, corresponding to  ${\cal L} = 85$, 12,
1.8, 0.11, respectively (compare with Fig. 1 in Park 2009).
Note that the rotation parameter ${\cal R}$ is small for all the
solutions, so these truly represent slowly-rotating flows.  Even the
most rapidly rotating solution (${\cal R}=0.31$) has a centrifugal
support of only 10\% of Keplerian at $r=r_{\rm B}$.

\begin{figure}
  \centering
  \includegraphics[width=0.97\columnwidth,angle=0]{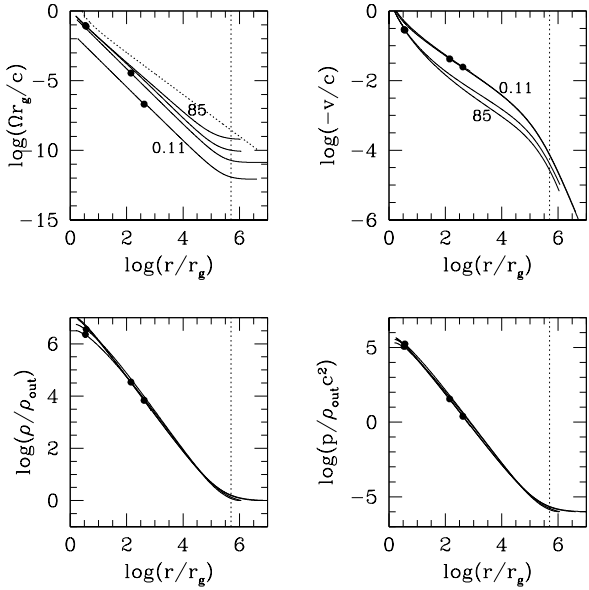}
  \caption{Representative solutions of the model equations for
  $\alpha=0.1$, $\gamma=5/3$, $c_{\rm out}=10^{-3}c$. The four
  solutions shown have $(\Omega_{\rm out}, ~{\cal R}, ~{\cal L},
  ~r_s)=$ ~$(0.624\times 10^{-9}$, 0.31, 85, 3.436); ~($0.851\times
  10^{-10}$, 0.043, 12, 3.663); ~($0.132\times 10^{-10}$, 0.0066, 1.8,
  142.0); ~($0.831\times 10^{-12}$, 0.00042, 0.11, 416.7),
  respectively. The solid dots indicate the positions of the sonic
  radii and are helpful for identifying the solutions.  In addition, a
  few curves are labeled by their values of $\cal{L}$. The vertical
  dotted lines correspond to the location of the Bondi radius $r_{\rm
  B}$ (eq. \ref{rbondi}), and the sloping dotted line in the top left
  panel shows the Keplerian angular frequency $\Omega_K$.}
\end{figure}

The solution with ${\cal L}=0.11$ -- the lowest curve in the top-left
panel of Fig. 1 -- is clearly in the Bondi regime since the gas has
negligible outer specific angular momentum relative to $l_{\rm ms}$.
The sonic radius $r_s$, shown by the black dot, is located at
$417r_g$, which is almost exactly where a pure non-rotating Bondi flow
has its sonic radius for our choice of $\Phi(r)$, $c_{\rm out}$ and
$\gamma$.  The two solutions with ${\cal L}= 85$ and 12 (the highest
two curves) are definitely rotation-dominated. The gas in these
solutions has too much angular momentum to permit steady accretion in
the absence of viscosity, so the accretion flow settles down to a
viscously-driven ADAF solution. Correspondingly, the sonic radius is
close to the marginally stable orbit, $r_{\rm ms}=3r_g$. The solution
with ${\cal L}=1.8$ represents a transition state between the Bondi
and ADAF regimes. Its sonic radius is at an intermediate location,
$r_s=142r_g$.

In Figure 2, the top-left panel shows how the sonic radius moves as we
change ${\cal L}$.  For all values of ${\cal L}<1$, $r_s$ is located
at the position one would calculate for the non-rotating Bondi problem
(upper dotted line), while for ${\cal L}$ greater than a few, $r_s$ is
close to $r_{\rm ms}$ (lower dotted line). The transition between
these two regimes is quite sudden, with most of the change happening
over the range $1.5 < {\cal L} < 2$.

\begin{figure}
  \centering
  \includegraphics[width=0.97\columnwidth,angle=0]{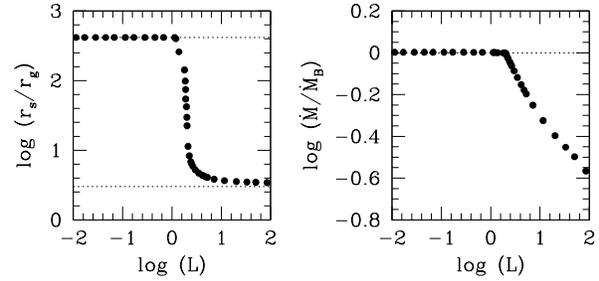}
\vskip -4.cm
  \caption{Left: Shows the location of the sonic radius $r_s$ as a
  function of the angular momentum parameter $\cal L$ for solutions
  with $\alpha=0.1$, $\gamma=5/3$, $c_{\rm out}=10^{-3}c$. The upper
  dotted line indicates the sonic radius for a pure non-rotating Bondi
  solution, and the lower dotted line shows the radius of the
  marginally stable orbit $r_{\rm ms}$. Note the sudden transition
  from a Bondi-like flow for ${\cal L} < 1$ to a rotation-supported
  ADAF for ${\cal L} > 2$.  Right: Shows the corresponding mass
  accretion rates $\dot{M}$ in units of the Bondi accretion rate
  $\dot{M}_{\rm B}$. The mass accretion rate is only a factor of three
  smaller than the Bondi rate even when $\cal L$ is as large as
  $\sim10^2$.}
\end{figure}

The bottom two panels in Fig. 1 show the profiles of density $\rho$
and pressure $p=\rho c_s^2$ for the same four solutions as in the top
left panel. Even though the rotation profiles of these solutions are
very different, and their sonic radii move around considerably, the
profiles of $\rho$ and $p$ are nearly identical. The insensitivity to
the location of $r_s$ is at least in part because we selected
$\gamma=5/3$, which is known to be a critical value of the adiabatic
index both for the Bondi problem and for ADAFs. Nevertheless, it is
clear that in many respects, an ADAF is very similar to a Bondi flow.

The top-right panel in Fig. 1 shows the radial velocity profiles of
the four solutions. We see that the radial velocity is smaller for a
rotating ADAF (the solutions with ${\cal L}=85$, 12) compared to a
slowly-rotating Bondi-like flow (${\cal L}=0.11$).  Since the density
profiles of both kinds of solution are nearly the same, this means
that the mass accretion rates are different. This is illustrated in
the panel on the right in Fig. 2 (compare with Fig. 2 in Park 2009),
which shows that $\dot{M}$ decreases as the rotation of the external
gas increases. The effect is quite modest, however --- the total range
of $\dot{M}$ in our solutions is only a factor of three, though this
is an artifact of our choice of a relatively large value of
$\alpha=0.1$.

To illustrate the effect of $\alpha$, Figures 3 and 4 show results
corresponding to ${\cal L}=13.5$ (${\cal R}=0.05$) and six values of
$\alpha$: 0.316, 0.1, 0.0316, 0.01, 0.00316, 0.001. The rotation
profiles are nearly the same for different values of $\alpha$, with
only small variations.  More interesting is the behavior of the sonic
radius, which moves in a very systematic way as $\alpha$ is varied.
For $\alpha=0.316$, we find $r_s=17.5$, which is well outside the
radius of the marginally stable orbit $r_{\rm ms}=3r_g$. With
decreasing $\alpha$, $r_s$ moves in until it is well inside $r_{\rm
ms}$.  The pattern is very similar to that seen in the ADAF models
described in Narayan et al. (1997) and noted in many other papers
(e.g., discussion of the slim disk model by Abramowicz et al. 2010).

The radial velocities shown in Fig. 3 decrease proportional to
$\alpha$, and so do the mass accretion rates (Fig. 4 right panel).
This is consistent with the predictions of the analytical ADAF model
described in Narayan \& Yi (1994), and is in qualitative agreement
with Park (2009). In particular, we agree with Park's conclusion that
low angular momentum flows resemble Bondi accretion, and that their
accretion rates approach the Bondi rate $\dot{M}_{\rm B}$ (which is
equal to $3.65\times10^8$ in our units where $r_g=c=\rho_{\rm out}=1$)
as the external rotation decreases.  However, we also see some
quantitative differences. Park suggests on the basis of his numerical
solutions that the accretion rate scales approximately as
$\dot{M}/\dot{M}_{\rm B}\sim 9\alpha/{\cal R}$ (our parameter ${\cal
R}$ is the same as $\lambda$ in Park's notation). We do not reproduce
the scaling with $\cal R$. For instance, Fig. 2 shows that, at fixed
$\alpha=0.1$, $\dot{M}$ changes by only a factor of $\sim3$ as ${\cal
R}$ changes by nearly a factor of 100. One reason for this difference
could be that we have considered solutions with $r_{\rm
B}/r_g=10^{5.7}$, whereas Park's solutions are closer to $10^3$. A
more thorough exploration of solutions in the three-dimensional space
of $\alpha$-${\cal R}$-$(r_{\rm B}/r_g)$ would be worthwhile to map
out how the accretion rate varies with these parameters.

The bottom two panels of Fig. 3 show that, with varying $\alpha$, the
density and pressure are largely independent of $\alpha$, just as we
earlier found them to be independent of $\Omega_{\rm out}$.  The
insensitivity of the central pressure to any parameter other than the
external density $\rho_{\rm out}$ and sound speed $c_{\rm out}$ is a
strong result and may have implications for jets (\S\ref{disc}).

\begin{figure}
  \centering
  \includegraphics[width=0.97\columnwidth,angle=0]{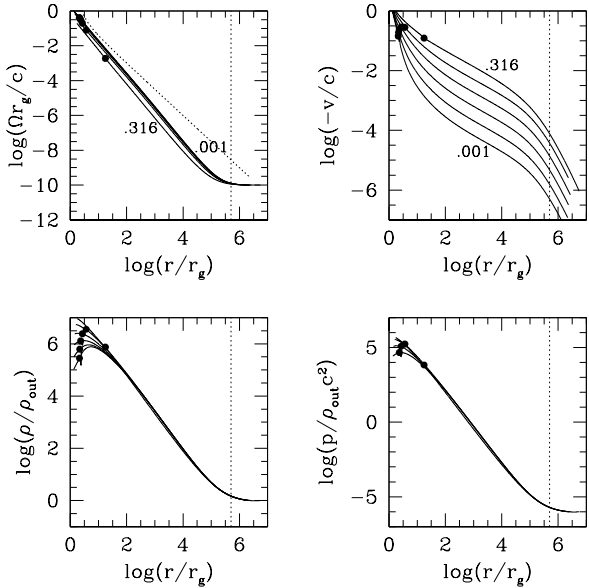}
  \caption{Solutions for $\gamma=5/3$, $c_{\rm out}=10^{-3}c$,
  $\Omega_{\rm out}=10^{-10}$, ${\cal R}=0.05$, ${\cal L}=13.5$, and
  six values of $\alpha$: 0.316, 0.1, 0.0316, 0.01, 0.00316,
  0.001. The six solutions have $r_s= 17.54$, 3.656, 2.694, 2.335,
  2.174, 2.114, respectively. The solid dots indicate the positions of
  the sonic radii and are helpful for identifying the solutions.  In
  addition, a few curves are labeled by their values of $\alpha$. The
  vertical dotted lines correspond to the location of the Bondi radius
  $r_{\rm B}$, and the sloping dotted line in the top left panel shows
  the Keplerian angular frequency $\Omega_K$.}
\end{figure}

\begin{figure}
  \centering
  \includegraphics[width=0.97\columnwidth,angle=0]{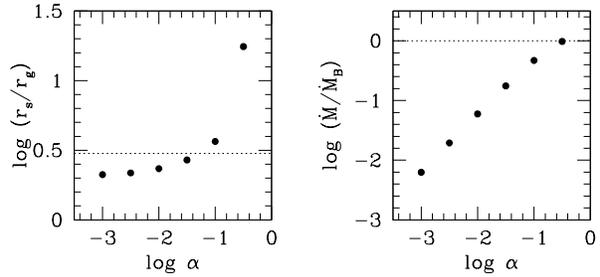}
\vskip -4.cm
  \caption{Left: Shows the location of the sonic radius $r_s$ as a
  function of the viscosity parameter $\alpha$ for the solutions
  described in Fig. 3.  The dotted line indicates the radius of the
  marginally stable orbit $r_{\rm ms}$. Large values of $\alpha$ cause
  the sonic radius to move outside $r_{\rm ms}$ while small values of
  $\alpha$ have the opposite effect.  Right: Shows the corresponding
  mass accretion rates $\dot{M}$ in units of the Bondi accretion rate
  $\dot{M}_{\rm B}$. Note that $\dot{M}$ is a steep function of
  $\alpha$, varying almost linearly. This is expected for an ADAF.}
\end{figure}

\section{Discussion}\label{disc}

The primary goal of this paper was to estimate the rate at which mass
accretes on to a supermassive black hole from rotating gas in the
nucleus of a galaxy.  We find that the answer depends on both the
dimensionless rotation parameter of the gas ${\cal R}$
(eq. \ref{rotpar}) and the viscosity parameter $\alpha$
(eq. \ref{visc}). For fixed $\alpha=0.1$, the accretion rate is within
a factor of a few of the Bondi rate for all values of ${\cal R}$
(Fig. 2), i.e., for this value of $\alpha$ accretion is nearly as
efficient in the presence of rotation as in its absence (the classic
Bondi problem).  Figure 4 shows the variation of $\dot{M}$ for fixed
${\cal R}=0.05$ (a reasonable value, see Inogamov \& Sunyaev 2010) and
different values of $\alpha$.  Here the variation is much larger.  The
accretion rate is suppressed by a large factor when $\alpha \ll 1$.
Hence, the answer to our primary question depends very much on the
value of $\alpha$.

King, Pringle \& Livio (2007) have examined a variety of observational
evidence and conclude that many observed accretion disks have $\alpha
\sim0.1-0.4$.  In addition, Sharma et al. (2006) argue that ADAFs have, 
if at all, even larger values of $\alpha$ compared to standard thin
disks.  Thus, the models with $\alpha=0.316$ and 0.1 in Fig. 4 may be
the best match to real radiatively-inefficient accretion flows in
galactic nuclei. If so, we can expect viscous accretion via an ADAF to
be quite efficient in galactic nuclei: $\dot{M}_{\rm ADAF}\sim
(0.3-1)\times\dot{M}_{\rm B}$.

As discussed in \S\ref{intro}, our picture is that a fraction
$\eta_{\rm acc}$ of the accretion energy near the black hole is
somehow converted to jet mechanical energy. If the accretion rate is
equal to the Bondi rate, then observations require about 2 per cent of
$\dot{M}_{\rm B}c^2$ to be transferred to the jet.  From the estimates
given above, we see that accretion of rotating gas via an ADAF
requires an efficiency of perhaps $\eta_{\rm acc} \sim 5\%$.  While we
do not have a model of how the jet is actually launched, an efficiency
of 5\% does not appear implausible.

It should be emphasized that the present study differs from most
previous discussions of this problem in the literature in that we
consider a steady viscous flow extending from beyond the Bondi radius
down to the black hole horizon. Viscosity enables our solutions to
overcome the centrifugal barrier and to accrete steadily, just as in
the standard thin accretion disk model (Shakura
\& Sunyaev 1973; Frank, King \& Raine 2002).  In contrast, most other
studies of quasi-spherical accretion with rotation (except Park 2009,
see \S\ref{intro}) have considered inviscid accretion. Those results
depend critically on the assumption that accretion is arrested once
the gas hits the centrifugal barrier.

From our numerical solutions, we can calculate the time required for
gas to travel from the Bondi radius down to the black hole.  For the
two models with $\alpha=0.316$ and 0.1 in Fig. 4, the accretion time
$t_{\rm ADAF}$ for the ADAF solution is no more than twice as long as
the accretion time $t_{\rm B}$ in the non-rotating Bondi solution.
Even for the rapidly rotating solution with ${\cal R}=0.31$ in Fig. 2,
$t_{\rm ADAF}$ is only $\sim 3t_{\rm B}$.\footnote{For very small
values of $\alpha$, we do we find $t_{\rm ADAF} \gg t_{\rm B}$, e.g.,
for $\alpha=0.001$, ${\cal R}=0.05$, we obtain $t_{\rm ADAF}/t_{\rm B}
= 156$. However, such low values of $\alpha$ seem unlikely.}  This is
very encouraging. For a turbulent external medium, we expect the
rotation of the external gas near the Bondi radius to vary on a time
scale of the order of tens of $t_{\rm B}$ (assuming a turbulent Mach
number $\sim 0.1$, Inogamov \& Sunyaev 2010). Since our ADAF solutions
have an accretion time much shorter than the turbulence time, there is
no problem setting up the steady state conditions we assume. More
importantly, the short accretion time guarantees that any feedback
from the ADAF via jets will occur rapidly compared to the dynamical
time of the external medium. Such instantaneous feedback is generally
assumed in most current models of feedback.

While we have focused so far on the accretion flow as the source of
jet power, a popular alternative hypothesis involves the black hole.
Blandford \& Znajek (1977) developed a scenario in which the
rotational energy of a spinning black hole is tapped by a magnetic
field and carried away in a magnetized jet. For a magnetic flux
$\Phi_B$ threading the horizon, Tchekhovskoy, Narayan \& McKinney
(2010) give a fairly accurate estimate of the jet power in geometrized
units ($GM=c=1$),
\begin{equation}
P_j = k \Phi_B^2 \Omega_H^2, \quad \Omega_H=\frac{a_*}{2r_H},
\quad r_H=2M[1+(1-a_*)^2]^{1/2},
\end{equation}
where $k\approx0.05$ is a dimensionless number that depends weakly on
the field geometry, $a_*\equiv a/M$ is the dimensionless spin
parameter of the black hole, and $r_H$ and $\Omega_H$ are the radius
and angular frequency of the black hole horizon.  The magnetic flux is
given by $\Phi_B=4\pi r_H^2 |B|_H$ where $|B|_H$ is the field strength
at the horizon. Thus
\begin{equation}
P_j = 32\pi^3k p_{\rm mag} r_H^2 a_*^2 \approx 50 p_{\rm mag} 
r_H^2 a_*^2,\label{Pjet}
\end{equation}
where $p_{\rm mag}=|B|_H^2/8\pi$ is the magnetic pressure at the
horizon.

To maintain a magnetic field on the horizon, it is necessary to keep
the field lines in place by means of an external pressure. We assume
that this pressure is supplied by the accretion flow.  Consider first
the Bondi non-rotating solution. In our units ($r_g=c=\rho_{\rm
  out}=1$), the thermal pressure of the Bondi solution at $r=2r_g$
is\footnote{We cannot use $r=r_g$ because our Newtonian potential
  (\ref{pw}) is singular there.}  $p_{\rm therm}(2r_g)=3.34\times
10^5$, and the ram pressure is $p_{\rm ram}(2r_g)=\rho v^2 =
6.51\times 10^6$, giving a total pressure of $p_{\rm
  tot}(2r_g)=6.85\times10^6$.  We assume that the total pressure is
what confines the central field and write the magnetic pressure at the
base of the jet as $P_B = \eta_B p_{\rm tot}(2r_g)$, where $\eta_B<1$
is a proportionality constant.  We also replace $r_H$ by $r_g$ in
equation (\ref{Pjet}), which is an overestimate for a rapidly spinning
hole but is probably a reasonable simplification. We then find
\begin{equation}
{\rm Bondi}: \qquad P_j \approx 3.4\times10^8\eta_B a_*^2
\approx \eta_B a_*^2 \dot{M}_B c^2.
\end{equation}
This interesting result shows that the jet power scales directly with
the Bondi accretion energy rate $\dot{M}_Bc^2$ and varies strongly
with the spin of the black hole. Our guess is that $\eta_B$ is
probably in the range $0.1-1$. Therefore, provided the black hole does
not rotate too slowly, a Bondi flow could easily support the jets seen
in observations.

We now compute the central pressures in the ADAF solutions.  The four
solutions shown in Fig. 2 have pressures ranging from $p_{\rm
  tot}(2r_g)=1.49\times10^6$ for ${\cal R}=0.31$ to $p_{\rm
  tot}(2r_g)=6.84\times10^6$ for ${\cal R}=0.00046$, while the
$\alpha=0.316$ and 0.1 solutions in Fig. 4 have $p_{\rm
  tot}(2r_g)=6.27\times10^6$ and $2.59\times10^6$, respectively.
These are the solutions most relevant for our problem, and their
pressures are between 20\% and 100\% of the Bondi pressure.  Thus, we
obtain the following estimate for the jet power in the presence of an
ADAF,
\begin{equation}
{\rm ADAF}: \qquad P_j \approx (0.2-1)\times \eta_B a_*^2 \dot{M}_B c^2.
\end{equation}
The jet efficiency factor in this model is
$\eta_j=(0.2-1)\times\eta_B$ and a net efficiency of 2\% seems quite
plausible.

To summarize, in terms of energetics at least, we have two viable
mechanisms to power jets via accretion in galactic nuclei: (i) by
tapping a fraction of the accretion energy, and (ii) by confining a
strong magnetic field around a spinning black hole and extracting
energy from the hole. Neither mechanism requires us to postulate
extreme conditions or to stretch parameters.  However, all of our
results are based on the simple one-dimensional model described in
\S\ref{ADAF}. Unfortunately, there are several important caveats that
need to be discussed.

Narayan \& Yi (1994, 1995a) showed that the accreting gas in an ADAF
has a positive Bernoulli parameter ${\cal B}$ (see eq. \ref{bernoulli}
for the definition). This means that the gas is gravitationally
unbound, and so these authors suggested that ADAFs would have strong
outflows and jets. The mass conservation equation (\ref{mass})
explicitly ignores such outflows.  How strong the outflows are is
difficult to estimate from first principles, though Blandford \&
Begelman (1999) suggested that the effect may be so strong that
$\dot{M}_{\rm ADAF}$ at the black hole might be reduced by orders of
magnitude.

Numerical hydrodynamic simulations have confirmed that the mass
accretion rate is indeed suppressed (Stone, Pringle \& Begelman 1999;
Igumenshchev \& Abramowicz 1999, 2000; Igumenshchev 2000), though
there is no consensus on whether gas truly escapes to infinity, or if
the Bernoulli parameter is even relevant (Abramowicz, Lasota \&
Igumenshchev 2000).  If outflows are as strong as Blandford \&
Begelman (1999) suggest, both jet mechanisms we have discussed here
are impossible. One mechanism depends directly on $\dot{M}_{\rm
ADAF}$, while the other depends on the central pressure $p_{\rm tot}$
which is roughly proportional to $\dot{M}_{\rm ADAF}$. If there are
heavy outflows, there is just not enough energy to power the observed
jets.

\begin{figure}
  \centering
  \includegraphics[width=0.97\columnwidth,angle=0]{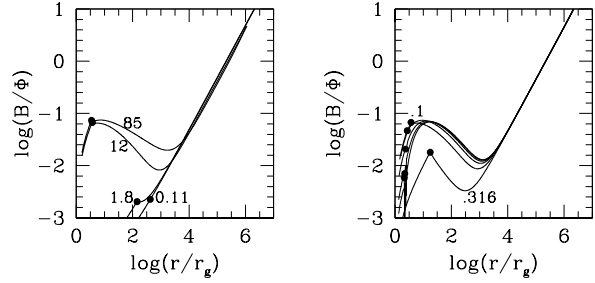}
\vskip -4.cm
  \caption{Variation with radius of the Bernoulli parameter $\cal B$
(eq. \ref{bernoulli}), normalized by the gravitational potential, for
the four solutions shown in Fig. 1 (left panel) and the six solutions
shown in Fig. 3 (right panel). The solid dots indicate the positions
of the sonic radii. Curves are labeled by their values of $\cal L$ in
the left panel and $\alpha$ in the right panel.}
\end{figure}

Figure 5 shows the variation of $\cal B$ with $r$ for the solutions
shown in Figs. 1 and 3. We have normalized ${\cal B}$ by the local
gravitational potential $|\Phi(r)|$ to obtain a dimensionless measure
of how unbound the gas is at each radius. All the solutions start with
a large value of ${\cal B}/|\Phi|$ outside the Bondi radius, but this
just means that the gas out there has a finite thermal energy but very
little gravitational binding energy. As the gas flows in, ${\cal B}$
hardly changes while $\Phi$ increases rapidly in magnitude, so the
ratio ${\cal B}/|\Phi|$ decreases rapidly. In the case of the
Bondi-like solution with ${\cal L}=0.11$, the decrease continues all
the way down to the horizon and there is no tendency to form an
outflow. The ADAF solutions, on the other hand, follow the Bondi
solution down to $r\sim10^3r_g$, after which viscous dissipation
causes ${\cal B}/|\Phi|$ to increase up to a maximum value of about
0.1.  We do not know if this value is large enough to strongly
suppress the mass accretion rate on to the black hole. However, since
the Bernoulli parameter is large only over a limited range of radius,
it is conceivable that outflows reduce the mass accretion rate in the
ADAF solutions by no more than a factor of a few, rather than by
orders of magntiude. In this case, both of our jet mechanisms are
likely to survive.

Our model assumes a single point source of gravitation, whereas a
realistic galactic nucleus has significant stellar mass from the inner
galaxy and any nuclear star cluster. This will make the outer gas more
gravitationallly bound than appears from Fig.~5. Radiative cooling
(which is ignored in our model) can also have an effect.  Using the
gas properties at the Bondi radius given in Allen et al. (2006) for
M87 we find that the radiative cooling time is about 2.5 percent of
the flow time ($r/c_{\rm s}$). The ratio of these timescales increases
at smaller radii, making radiative cooling unimportant there. Compton
cooling however increases with decreasing radius and may be important
at precisely those inner radii where $\cal B$ shows an increase in
Fig. 5.  Thus, the importance of the Bernoulli parameter may be
reduced even further.

Another important effect pointed out by Narayan \& Yi (1994) is that
viscous dissipation, coupled with the lack of radiative cooling,
causes the entropy of the gas in an ADAF to increase inward, making
the flow convectively unstable by the Schwarzschild criterion.
Convective effects have not been included in the one-dimensional model
we have considered in this paper.  Narayan, Igumenshchev \& Abramowicz
(2000) and Quataert \& Gruzinov (2000) discussed the physics of
convection-dominated accretion flows (CDAFs) and concluded that such
flows would differ enormously from ADAFs.  In particular, if one
considers a self-similar model, the density, pressure and mass
accretion rate of a CDAF, as measured at the black hole, are predicted
to be a factor $\sim r_g/r_B \sim 10^{-5}$ (for our models) times the
corresponding values for an ADAF with the same outer boundary
conditions.  Even if the real effect is only a small fraction of this
analytical prediction, it would reduce jet power to a level far below
that observed.

Numerical hydrodynamic simulations confirm the presence of convection
in ADAFs (Stone et al. 1999; Igumenshchev \& Abramowicz 1999, 2000;
Igumenshchev, Abramowicz \& Narayan 2000) and indicate substantial
suppression of the mass accretion rate into the black hole. This is
problematic for jet production.  However, the relative importance of
ADAFs vs CDAFs has been debated (e.g., Igumenshchev \& Abramowicz
1999; Abramowicz et al. 2002; Lu, Li \& Gu 2004), and the issue is
still unresolved.

\begin{figure}
  \centering
  \includegraphics[width=0.97\columnwidth,angle=0]{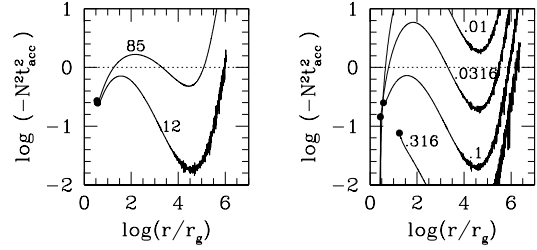}
\vskip -4.cm
  \caption{Left: Variation with radius of the quantity $-N^2 t_{\rm
  acc}^2$, which is a measure of convective instability, for two of
  the four solutions shown in Fig. 1. (The other two are below the
  bottom of the plot.) The dotted line indicates our nominal threshold
  for becoming convection-dominated. The more rapidly spinning
  solution (${\cal L}=85$) goes above the dotted line over a range of
  radius, and we expect it to become a CDAF in this zone. The less
  rapidly rotating solution (${\cal L}=12$) lies below the dotted line
  and probably will not become a CDAF.  Right: Corresponding results
  for the solutions shown in Fig. 3. The two solutions with
  $\alpha=0.316$ and 0.1 lie below the dotted line and will not be
  convection-dominated, while the solutions with smaller values of
  $\alpha$ are expected to become CDAFs.  Note that none of the curves
  in the two panels extends inside the sonic radius. This is because
  we have assumed the flow to be inviscid once it becomes
  supersonic. The specific entropy is then constant
  (\S\ref{supersonic}) and $N^2=0$.}
\end{figure}

A measure of convective instability is the Brunt-Vaisala frequency $N$
which for a spherically symmetric system is given by
\begin{equation}
N^2 =-\frac{1}{\rho\,}\frac{dp}{dr}\,\frac{d\ln(p^{1/\gamma}/\rho)}{dr}.
\label{brunt}
\end{equation}
A system is convectively unstable if $N^2<0$. When this happens, the
quantity $|N|\equiv \sqrt{-N^2}$ measures the growth rate of the
instability. We imagine that convection is important and takes over
the dynamics of the flow only when the growth time scale of the
instability is shorter than the accretion time $t_{\rm acc}\equiv
r/|v|$, whereas in the opposite limit, we expect convection to be a
minor perturbation. Motivated by this argument, we show in Fig. 6 the
profiles of the dimensionless quantity $-N^2t_{\rm acc}^2$ versus $r$
for a selection of our numerical solutions.  Most of the solutions of
interest to us, viz., those with relatively large values of $\alpha$
and small rotation parameters $\cal R$, have $-N^2t_{\rm acc}^2 < 1$.
Convection is probably unimportant in these cases.  The results shown
in the panel on the right agree with the $\alpha$ trend discussed by
Igumenshchev \& Abramowicz (1999) and Lu et al. (2004).

In the discussion so far, we have ignored the effect of magnetic
fields. Magnetohydrodynamic (MHD) simulations of ADAFs (Stone \&
Pringle 2001; Hawley, Balbus \& Stone 2001; Igumenshchev \& Narayan
2002; Igumenshchev, Narayan \& Abramowicz 2003; Pen, Matzner \& Wong
2003; Igumenshchev 2006, 2008; Pang et al.  2010) have not so far
exhibited strong unbound winds, but they nevertheless have shallow
density profiles and reduced $\dot{M}$.  The flows exhibit vigorous
turbulence and they transport energy outward, just as one expects with
a convective flow. However, whether or not the turbulence can be
described as convection is unclear (Narayan et al. 2002, Balbus \&
Hawley 2002; Pen et al. 2003).

From the point of view of our present study, the relevant question is
how much is the mass accretion rate suppressed relative to the Bondi
rate as a result of MHD effects.  The best scalings from current
simulations suggest that the accretion rate, and therefore the jet
power, is reduced by at least three orders of magnitude.  Moreover,
$\dot{M}$ is found to be reduced substantially even in the case of a
non-rotating Bondi flow (Igumenshchev \& Narayan 2002; Igumenshchev
2006). This last result is particularly worrisome in view of the
results shown in Fig. 6. There it appeared that, so long as $\alpha$
is relatively large and the rotation is small, hydrodynamic convection
and the consequent suppression of $\dot{M}$ are not an issue. However,
it appears that magnetic fields completely alter the situation and
strongly suppress even Bondi accretion. This effect needs to be
confirmed with further studies and the physics of the phenomenon needs
to be identified.

Apart from convection, thermal conduction might also modify the
dynamics of radiatively inefficient accretion (e.g., Johnson \&
Quataert 2007; Sharma, Quataert \& Stone 2008; Shcherbakov
\& Baganoff 2010). Numerical MHD simulations including conduction and
covering an adequate range of radius are yet to be carried out.

If any of the effects discussed here succeeds in strongly suppressing
the mass accretion rate in quasi-spherical accretion flows, we would
be left with the puzzle of why the observed jet power $P_j$ in many
nearby low-luminosity galactic nuclei tracks the estimated Bondi mass
accretion rate $\dot{M}_{\rm B}$ (\S\ref{intro}): $\eta_j\equiv
P_j/\dot{M}_{\rm B}c^2\sim 2\%$. The observed jets would require a
different explanation that is not related to a hot accretion flow, and
the apparent correlation with properties of hot gas at the Bondi
radius must be a coincidence.

A disk of dense cool gas emitting optical lines is often seen in the
nuclei of elliptical galaxies. Macchetto et al. (1997) have studied
the gas distribution in M87 and used the kinematics of the gas to
estimate the mass of the black hole. The result is slightly less than
that now determined from stellar kinematics by Gebhardt \& Thomas
(2009). The gas disk has a central hole a few pc in size. Line widths
are large, possibly indicating the action of nongravitational forces
in this and other elliptical galaxies (Verdoes-Kleijn, van der Marel
\& Noel-Storr 2006). We do not think that small masses of such cool
gas are incompatible with the existence of a giant ADAF.

Finally, we note that estimates showing that Bondi accretion cannot
yield a sufficient mass accretion rate (e.g., Rafferty et al. 2006)
are based on the temperature and density values inferred at radii far
outside $r_{\rm B}$. Small quantities of cooler gas at the centre, say
at $0.7\keV$ instead of $3\keV$, which is in pressure equilibrium with
the surrounding gas would have a correspondingly higher density and
allow a much higher Bondi flow rate ($\dot M_{\rm B}
\propto T^{-5/2}$). Such cool gas is often observed in nearby clusters
where the innermost regions are spatially-resolved.  The rate will in
practice be even higher than this simple scaling as the pressure will
be higher at the Bondi radius due to the weight of intervening gas.

\section{Summary}

We have shown in this paper that accretion can occur from a hot
atmosphere at close to the Bondi rate, provided the external gas near
the Bondi radius rotates relatively slowly (less than a few tens of
percent of the Keplerian rate) and the viscosity parameter $\alpha$ is
fairly large ($\geq 0.1$). The non-radiative numerical ADAF solutions
computed here may be relevant to the nuclei of massive elliptical
galaxies hosting a central black hole surrounded by a hot gaseous
atmosphere. The mass accretion rate is large enough that it requires
only a small fraction ($\sim2-5\%$) of the accretion energy to power
the observed jets in nearby elliptical galaxies.  At the same time,
the jets can heat the surrounding gas and prevent the hot atmosphere
from radiatively cooling and collapsing into the centre. This feedback
mechanism could also have an effect on the evolution of the galaxy
itself. These results require that mass outflows (other than that
associated with the jets), convective energy transport and MHD effects
are weaker than usually assumed.

\section*{Acknowledgements}
We thank the referee for a thoughtful review. RN thanks the Institute 
of Astronomy, Cambridge, for hospitality while
much of this work was carried out.  RN's research was supported in
part by NSF grant AST-1041590 and NASA grant NNX11AE16G.  ACF thanks
the Royal Society for support.

\bibliographystyle{mnras}

\end{document}